\documentclass{JHEP}
\usepackage{amsmath}
\usepackage{amssymb,amsfonts}
\usepackage{epsfig}

\advance\hoffset by -.35in
\newcommand{\ex}[1]{{\rm e}^{#1}}
\preprint{hep-th/0012211\\DAMTP-2000-140}

\title{\Large\bf Taming the supergravity description of non-BPS
D-branes~: the D{/}\={D} solution.}

\author{P. Bain\\

\vskip 24pt

Department of Applied Mathematics and Theoretical Physics,\\
University of Cambridge,\\
Wilberforce Road, CB3 0WA Cambridge, U.K.\\
E-mail: \email{P.A.Bain@damtp.cam.ac.uk}}

\abstract{We obtain the supergravity solution which describes a bound
state of D-string{/}anti-D-string pairs attached to different fixed
planes of an orbifold, in type IIB string theory compactified on
$T^4/{\mathbf 
Z}_2$. For parameters at which the conformal field theory
point of view predicts stability, the solution displays a repulson-like
singularity. However, we observe that a D-string{/}anti-D-string
pair probe in this background becomes tensionless before reaching the
singularity, suggesting a resolution by the enhan\c{c}on
mechanism. Moreover, the force feels by this probe is attractive,
in contrast to the repulsive behaviour observed in the 
non-BPS D-brane description.}


\begin{document}
\section{Introduction}

Non-BPS D-branes have received a lot of attention in the last couple of
years as a way to test duality symmetries between string theories
beyond the supersymmetric spectrum \cite{sen1, sen2, sen3, sen4, sen5,
senr, leru, schwarz, gaberdiel}. 
These studies mainly rely on the conformal field theory
description of the non-BPS branes, as dynamical surfaces on which open
strings end \cite{polchinski}. However, we know that BPS-branes also
enjoy a  
space-time description as classical solutions of the field equations
of supergravity \cite{duff}. While the conformal field theory 
description is valid at weak coupling, the supergravity solution is a
good description of a strong interacting system, especially for a
large number of D-branes. 

A crucial ingredient for
superposing an arbitrary number of BPS D-branes is the so-called
``no-force'' condition, which is a consequence of the presence of
preserved supersymmetries in
this context~: since these D-branes do not exercise any force
on each other, one can bring $N$ of them from infinity to form a
macroscopic bound state which has a reliable classical description. 

On the other hand, in general,  stable non-BPS D-branes do not obey
such a constraint.  Actually, only non-BPS D-branes of orbifold
compactification 
of type II string theories have been found to enjoy this property at
one-loop \cite{gabsen}. Indeed, for specific radii of
compactification, a Bose-Fermi 
degeneracy appears and ensures that the no-force condition is satisfied
at this order in the coupling constant. 
Therefore, assuming that such D-branes should have a classical
description in supergravity, several papers tried to find the explicit
solution \cite{eyras, loz, lerda} (see also \cite{brax} for a study in
the ten dimensional context). They found that, as expected, the
solution verifies the 
no-force condition at one-loop for a critical volume but, at higher
orders, the potential 
felt by a non-BPS D-brane probe in the background geometry created by
the other D-branes is not constant~: the no-force
condition fails and these non-BPS branes actually exercise a repulsion
on each other at any distance, which prevents them to form a bound state
\cite{lerda}. 
Moreover, the metric is plagued with naked singularities, located at
finite proper distances from the core of the brane, and no 
mechanism to hide them has been found so far.

More recently, Lambert and Sachs \cite{lambert} argued that, at 
critical radii, the
non-BPS D-brane states considered 
in \cite{eyras, lerda} were not the minimum of the tachyon
potential. The 
true vacuum is obtained when the world-volume scalars associated to
the tachyon winding modes condense; in the type II string theory
compactified on $T^4/{\mathbf Z}_2$ picture, it is described by a system of
D-brane{/}anti-D-brane located at two different fixed points of the
orbifold, as we will review in detail in the next section. They
calculated the one-loop potential and they found that, in this phase, two
non-BPS branes attract  each other, independently of the K3
volume. Therefore, it seems to be a good 
starting point to find a classical description. 
The aim of the present article is to provide such a solution for the
D-string{/}anti-D-string system in type IIB theory on $T^4/{\mathbf
Z_2}$. However, the calculation could be extended easily to other
D$p$-branes 
configurations, at least for $p$ less than three. Indeed, for $p=3$, we
do not expect 
anymore an asymptotically flat space, since the supergravity fields will
have a logarithmic dependence on the radial coordinate. 

This paper is organized as follows. In the next section, we 
review the conformal field theory description of non-BPS D-branes in
type IIB string theory on $T^4/{\mathbf Z_2}$. In particular, we 
give their associated boundary states which will be useful to obtain
the boundary action and the 
asymptotic behaviour of the supergravity solution. We also review
the boundary state obtained at the minimum of the tachyon
potential, as discussed in \cite{lambert}. 
In section three, we write the six-dimensional supergravity lagrangian
which describes the dynamics of the fields excited by the D-brane
configurations. We also discuss the boundary actions which reproduce
the linear couplings of the supergravity fields to the non-BPS
D-branes. 
Another useful information given in this section is the asymptotic
behaviour of the bulk fields in 
the presence of a non-BPS D-brane or of a D-string$/\bar{\rm
D}$-string pair. The subject of this section has a significant overlap
with  
\cite{frau}, which appears while our work was in progress. Therefore,
to avoid repetitions, we will borrow its notations and results 
and extend them to the non-BPS configurations.  

In section four, we first discuss the issue of the no-force condition
in our configuration and then solve the equations of motion. Details
are given in an appendix. We observe that the solution, which depends
on the moduli of K3 only through its volume,  always displays
naked singularities. However, by studying the behaviour of a slowly
moving 
probe in this background, we see that the brane becomes tensionless
before reaching the first singularity. Therefore, we argue that an 
enhan\c{c}on mechanism similar to the one advocated in \cite{john}
should provide a resolution of the singularities.  In particular, we
show that, on the tensionless sphere, new massless bulk states are
present and the gauge symmetry is enhanced. 
Finally, the last section contains some comments and discusses a related
problem.

%%%%%%%%%%%%%%%%%%%%%%%%%%%%%%%%%%%%%%%%%%%%%%%%%%%%%%%%%%%%%%%%%%%%%%%%%%
\section{The non-BPS D2-brane and the D-string{/}anti-D-string system
in type 
IIB string theory on $T^4/{\mathbf Z_2}$}

\subsection{The supergravity}

The context of our study is type IIA string theory compactified
on $T^4/(-1)^{F_L}{\cal I}_4$ where ${F_L}$ is the left-mover
spacetime fermion number. By T-duality along one of the orbifold
directions, this theory is related to the type IIB string theory on
$T^4/{\cal I}_4$.  In this paper, we will not try to find supergravity
solutions in ten dimensions, with D-branes localized at some fixed points
of the orbifold. Indeed, 
such study would involve the construction of dipoles of (fractional)
D-branes, dipoles which are not yet well understood, even in the
context of 
standard BPS D-branes in ten dimensions~: so far, only dipoles solutions
depending of three transverse directions have been discussed in the
literature \cite{youm, janssen, emparan, chatta}. Moreover, the
supergravity description of 
fractional D-branes has appeared only recently, first for D3-branes on
a non-compact ${\mathbf Z}_2$ orbifold \cite{bertolini} and then, for
D0-branes in type 
IIA string theory compactified on $T^4/{\mathbf Z}_2$ \cite{frau}. 

The six-dimensional low energy effective theory is the  
chiral ${\cal N}_6=(2,0)$ supergravity and
contains the graviton multiplet and twenty-one tensor multiplets. 
From the type 
IIB point-of-view, the bosonic content of these multiplets comes from
the reduction of the ten-dimensional NS-NS and R-R fields on K3. The
reduction of the metric gives the six-dimensional metric tensor $g$
and 58 scalars. In the following, we will be particularly interested
into the four scalars which correspond to the diagonal elements of the
metric along the compact directions and which will be denoted $\eta^a$,
as in \cite{lerda, frau}.  
The reduction of the NS-NS 2-form on the supersymmetric 2-cycles of
K3 gives 22 scalars, which can be decomposed into 6 untwisted states
associated to the 2-cycles of the 4-torus and 16 twisted moduli,
called $b^I \, (I=1, \ldots, 16)$ below and which are 
associated to the 16 exceptional supersymmetric 2-cycles of K3. In
the orbifold limit, these twisted fields are localized at the 16 fixed
points of the orbifold. 
From the R-R sector, one obtains R-R twisted 2-forms and twisted
scalars. Explicitly, the ten dimensional R-R 4 and 2-form, $C_{(4)}$
and $C_{(2)}$, decompose as~:
\begin{eqnarray}
C_{(4)} &=& C^6_{(4)} + \sum_{i=1}^6 C^i_{(2)} \wedge \omega^i_2 +
\sum_{i=1}^{16} {\cal A}^I_{(2)} \wedge \tilde{\omega}^I_2 + \tilde{C}_{(0)} \;
\omega_4 \nonumber \\ 
C_{(2)} &=& C^6_{(2)} + \sum_{i=1}^6 C^i_{(0)} \; \omega^i_2 +
\sum_{i=1}^{16} {\cal A}^I_{(0)} \; \tilde{\omega}^I_2,
\label{expansion}
\end{eqnarray}
where $\omega^i_2$, $\tilde{\omega}^I_2$ and $\omega_4$ denote
respectively the harmonic forms dual to the six 2-cycles of the 4-torus,
to 
the sixteen vanishing 2-cycles and to the 4-cycle. Notice that
the 4-form $C^6_{(4)}$ can be Hodge-dualized to a scalar. Taking into
account the anti-self-duality of the exceptional 2-cycles 
$\tilde{\omega}_2^I$, the ten  dimensional self-duality of the field
strength of $C_{(4)}$ translates into   
anti-self-duality conditions for the field strengths of the 2-forms
${\cal A}^I_{(2)}$.  In the
following, we will forget the ``6'' index since we will work exclusively
in six dimensions.

Contrary to the non-chiral theory considered in
\cite{lerda}, the six-dimensional lagrangian which describes the
dynamics of this chiral ${\cal N}_6=(2, 0)$ supergravity can not be
obtained from the Kaluza-Klein reduction of a string theory on a
4-torus. Therefore, we will adopt another method, also used in
\cite{frau} in the context of type IIA string theory on $T^4/{\mathbf
Z}_2$. Moreover, we will not write the full lagrangian, but only the
terms which are relevant for our study, {\it i.e.} the bulk fields excited by
the D-string{/}anti-D-string sources. To select 
these fields, we first have to describe precisely the sources, what
we will do in the next two subsections.

\subsection{The non-BPS D-brane}

We will limit our study to the non-BPS D-string of type IIA compactified
on the orbifold $T^4/(-1)^{F_L}{\cal I}_4$. This D-string, which is
localized at one of the sixteen fixed points of the orbifold space,
say ${\mathbf x}_1 =(0, 0, 0, 0)$, is stable
provided that all radii, $R_a, a=6, \ldots, 9$, of the four-torus are
larger than $\sqrt{\alpha^\prime/2}$ \footnote{Latin indices $a, b,
\ldots$ will 
denote the four compact coordinates $x^6, \ldots, x^9$. We will also
use greek indices 
$\alpha, \beta, \ldots$ for the 
non-compact longitudinal coordinates and latin letters $i, j,
\ldots$ for the non-compact transverse  coordinates.}.
Indeed, the tachyonic ground state of
the NS sector of open strings is projected out by $(-1)^{F_L}{\cal
I}_4$ and only odd winding modes survive. Therefore, the square masses
of the four lightest states,
\begin{eqnarray*}
\frac{\chi^a}{2 i}\left(e^{i X^a R_a/\alpha^\prime}-e^{-i X^a
R_a/\alpha^\prime} \right), \qquad 
\alpha^\prime m_a^2 = - \frac{1}{2} + \frac{R^2_a}{\alpha^\prime},
\qquad a=6, \ldots, 9
\end{eqnarray*}
are positive for $R_a \geq \sqrt{\alpha^\prime/2}$. 

When one of these radii, for instance 
$R_9$, becomes smaller that this critical value, the field $\chi^9$ is
tachyonic; the non-BPS D-string is unstable and decays into a 
D2/\=D2-branes pair, stretched between the two fixed points along the
$x^9$ direction, with a ${\mathbf Z}_2$ Wilson line on one of the
D2-branes \cite{sen11, bergab}. Moreover, when all radii are critical, the
non-BPS D-string is stable and its spectrum has a Bose-Fermi degeneracy
\cite{gabsen}. Therefore, the one-loop open string partition function
vanishes and the no-force condition between two such non-BPS objects
is verified at this order. 

From the conformal field theory point-of-view, this non-BPS D-string is
described by a boundary state made up of an untwisted NS-NS part and
a twisted R-R part \cite{bergab}~:
\begin{equation}
|D1 \rangle = 
\frac{T_1}{\sqrt{2 V}}|B1, {\mathbf x}_1 \rangle_{\rm NS-NS}^U +
\frac{T_1}{\sqrt{2}\pi^2\alpha^\prime}|T_1 \rangle_{{\rm R-R}}.   
\label{nbps1}
\end{equation}
The explicit form of 
the boundary states $|B1, {\mathbf x}_1 \rangle_{\rm NS-NS}^{U}$ and $|T_1
\rangle_{{\rm R-R}}$ can be found for instance in \cite{frau}. 

In the T-dual type IIB language, this non-BPS state corresponds to a
non-BPS D2-brane stretched between the two fixed planes along the
T-dualized direction \cite{gaberdiel}. After T-duality, the new
stability constraint on the radii reads $R_a \geq
\sqrt{\alpha^\prime/2}$ for $a=6, 7, 8$ and $R_9 \leq \sqrt{2\alpha^\prime}$,
if $x^9$ is the T-dualized direction. When $R_9$ becomes larger than 
$\sqrt{2\alpha^\prime}$, the non-BPS D2-brane decays into a fractional 
D-string/anti-D-string pair, each localized at opposite fixed points,
${\mathbf x}_1 = (0, 0, 0, 0)$ and ${\mathbf x}_2 = (0, 0, 0, \pi
R_9)$ along the $x^9$ 
direction \cite{sen5, gaberdiel2}. The boundary state which
corresponds to this non-BPS D2-brane is given by~:
\begin{equation}
|\hat{D2} \rangle = 
\frac{T_1 R_9}{2 \sqrt{V \alpha^\prime}}|B2 \rangle_{\rm NS-NS}^U +
\frac{T_1}{2\sqrt{2}\pi^2\alpha^\prime}\left(|T_1 \rangle_{{\rm R-R}} +
|T_2 \rangle_{{\rm R-R}} \right). 
\label{nbps}
\end{equation}
Other decay channels, obtained when $R_a < \sqrt{\alpha^\prime/2}$, are
described in \cite{gaberdiel}.

Using these boundary states, one can first extract the linear couplings
of the bulk fields to the D-branes, by evaluating one-point
functions, and, second, find the asymptotic behaviour of these fields
at large distance from the D-branes \cite{bs}. 
The method has been used in the context of the non-BPS D-particle of
type IIB string theory compactified on $T^4/(-1)^{F_L}{\cal I}_4$ in
\cite{eyras} 
to obtain the large distance dependence of the bulk fields. Since
these calculations encode only the one-loop effect of string theory,
one recovers the no-force condition when the radii are critical.  

However, to find the complete solution and not only the asymptotic
behaviour, it is easier to solve directly the equations of motion with
the appropriated source terms rather than to calculate higher loops
diagrams. This program has been settled in \cite{lerda}, where they
considered a source made of a superposition of $N$ non-BPS D0-branes in
type IIB on $T^4/(-1)^{F_L}{\cal I}_4$ and they 
succeeded to  find the complete solution which described this state.  
Their solution, which is exact in the large t'Hooft coupling 
($Ng_s \rightarrow \infty$),  exhibits pathological behaviours~: first,
their solution displays naked repulson-like singularities which have
not been cured so far;  the enhan\c{c}on mechanism which has recently been
put forward to solve similar singularities in other contexts does not
seems to apply here since characteristic properties such as the
existence of a tensionless locus or the enhancement of gauge
symmetries are missing in their solution. Moreover the no-force
condition is not satisfied for higher string loops than one-loop;
actually, the force between the non-BPS branes is always repulsive, 
preventing them to form a localized bound state and, therefore,
contradicting the initial hypothesis.  This implies that one can not  
construct a reliable supergravity solution with small enough curvatures  
to neglect the $\alpha^\prime$ corrections to the bulk
supergravity. Although the result is disappointing from the point of
view of the 
quest of a classical description for non-BPS D-branes, the
supergravity approach is fruitful since it gives us insights into
loops corrections which are difficult to handle on the string
theory side.

\subsection{The D-string{/}anti-D-string configuration} 

An argument corroborating this discover came from the analysis of 
loops corrections to the tachyon potential \cite{lambert}. They
found that the non-BPS D-brane configuration 
considered in \cite{lerda} is not the minimum of the tachyon potential
when the radii are critical. Instead, they argued that the true vacuum 
is obtained by giving a non vanishing vacuum expectation value for one
of the 
four tachyon fields, say the state $\chi^9$ (we still call them
tachyon fields even if they 
are massless at critical radii). According to Sen \cite{sen11}, this
corresponds to a splitting of the non-BPS D-brane into the D/\=D pair
with a relative ${\mathbf Z}_2$ Wilson line in which the non-BPS
D-brane would decay when $R_9 > \sqrt{\alpha^\prime/2}$. To summarize,
the main result of \cite{lambert} was to show that, for critical radii,
while the analysis of the open string spectrum does not discriminate
between the non-BPS D-brane and the D/\=D phases, the
minimization of the tachyon potential selects the latter state. 

In the T-dual type IIB on $T^4/{\cal I}_4$ picture that we will use in
the rest of this article, this vacuum is given by
D-string{/}anti-D-string pairs sitting at opposite fixed points along
the compact coordinate, all D-branes at one point (${\mathbf x}_1$) and all
anti-D-branes at the other (${\mathbf x}_2$).  
Since we want to find the supergravity solution which describes this
vacuum, we will first review the boundary state associated to this
configuration. The (anti-)D-strings are fractional BPS (anti-)D-branes,
which can be interpreted as (anti-)D3-branes wrapped on vanishing
supersymmetric 2-cycles of the orbifold. The boundary state
for the D-string reads
\begin{eqnarray}
|D1 \rangle =&& \frac{T_1}{2\sqrt{2 V}} \left(|B1, {\mathbf x}_1 \rangle_{\rm
NS-NS}^U + |B1, {\mathbf x}_1 \rangle_{\rm 
R-R}^U\right) +
\frac{T_1}{2\sqrt{2}\pi^2\alpha^\prime}\left(|T_1\rangle_{\rm NS-NS} +
|T_1 \rangle_{\rm R-R}\right) 
\label{boundary-d} 
\end{eqnarray}
while, for the anti-D-string, it is given by
\begin{eqnarray}
|\bar{D}1 \rangle =&& \frac{T_1}{2\sqrt{2 V}} \left(|B1, {\mathbf x}_2
\rangle_{\rm NS-NS}^U - |B1, {\mathbf x}_2 \rangle_{\rm 
R-R}^U\right) - \frac{T_1}{2\sqrt{2}\pi^2\alpha^\prime}\left(|T_2
\rangle_{\rm NS-NS} - |T_2 \rangle_{\rm R-R}\right) 
\label{boundary-ad} 
\end{eqnarray}
where the explicit forms of the boundary states in the untwisted and
twisted sectors can be found in \cite{frau}. 

These D-strings have opposite charges under the untwisted R-R
2-form. Since we will only discuss in this article the
six-dimensional solutions, we will not see the dipole effect due to
the separation of the branes along the orbifold
directions. The boundary state we should consider to extract the 
couplings to 
the brane and the asymptotic behaviour of the supergravity fields is
obtained by ``neglecting'' the dependence on the orbifold coordinates
and summing these two boundary states; the untwisted
R-R contributions cancel and it remains only~:
\begin{eqnarray}
|D1; \bar{D}1 \rangle =&& \frac{T_1}{\sqrt{2 V}} |D1 \rangle_{\rm NS-NS}^U +
\frac{T_1}{2\sqrt{2}\pi^2\alpha^\prime} \left(|T_1 
\rangle_{\rm NS-NS} + |T_1 \rangle_{\rm R-R} - |T_2 \rangle_{\rm
NS-NS} + |T_2 \rangle_{\rm R-R} \right). \nonumber \\
\label{dd} 
\end{eqnarray}
Compared to the boundary state (\ref{nbps}), the key difference is that
the D-string{/}anti-D-string configuration also couples to the two twisted
NS-NS scalars $b^1$ and $b^2$ localized at the fixed points ${\mathbf
x}_1$ and 
${\mathbf x}_2$. From the boundary states, we can extract the couplings of the
D-branes to the closed string fields. However, we will refrain to give
them now and postpone this presentation after the description of the
six-dimensional supergravity action which governs the dynamics of
the bulk fields.

%%%%%%%%%%%%%%%%%%%%%%%%%%%%%%%%%%%%%%%%%%%%%%%%%%%%%%%%%%%%%%%%%%%%%%%%%%
\section{Low energy effective actions}

\subsection{The six-dimensional supergravity action}

To write the effective action which describes the dynamics of the
fields excited by the branes, we will start from the ten dimensional
action and perform a Kaluza-Klein reduction on the orbifold space
$T^2/{\mathbf Z}_2$. This approach was also  adopted in \cite{frau}
for the IIA string theory, where a consistency check can be made by
comparing with the action obtained through duality 
with heterotic string theory compactified on a 4-torus. Since type IIA
and type IIB strings share the NS-NS untwisted and twisted sectors,
we can read directly the corresponding kinetic terms in
\cite{frau}. The difference lies in the R-R part and the
Wess-Zumino term but these couplings can be easily extrapolated from
the knowledge of their type IIA counterparts. 

We  start with the ten dimensional action of type IIB string
theory, which, in the string frame, reads\footnote{The gravitational
coupling constant is given by $\sqrt{2}\kappa_{10}=(2\pi)^{7/2}
\alpha^{\prime 
2} g_s$. It includes the asymptotic value of the dilaton.}~:
\begin{eqnarray}
{\cal S}^{B,\,\sigma}_{10} &=& \frac{1}{2\kappa^2_{10}} \Bigg[\int d^{10}x
\sqrt{-{\det}_{10}{g}}\; e^{-2\phi} R - 
 \int  
e^{-2\phi}\; \left(4\;d\phi \wedge^\star\!{d\phi} + \frac{1}{2}
H_{(3)} \wedge^\star\!{H_{(3)}}\right) \nonumber
\\ 
&&\left.
- \frac{1}{2} \left( dC_{(0)} \wedge^\star\!{dC_{(0)}} + 
\tilde{F}_{(3)} \wedge^{\star}\!{\tilde{F}_{(3)}} +
\frac{1}{2} \tilde{F}_{(5)} \wedge^\star\!{\tilde{F}_{(5)}} + 
C_{(4)} \wedge H_{(3)} \wedge F_{(3)} \right)\right]
\label{action10d}
\end{eqnarray}
where the field strengths of the NS-NS 2-form and of the R-R 2- and
4-forms potential are defined as
\begin{eqnarray}
H_{(3)} = dB_{(2)},~~~~ F_{(p+1)} = dC_{(p)} ~~~~ \mbox{and} ~~~~
\tilde{F}_{(p+1)} = F_{(p+1)} - C_{(p-2)} \wedge H_{(3)}. \nonumber
\end{eqnarray}
The self-duality of $\tilde{F}_{(5)}$ is imposed at the level of the
equations of motion. 

As stressed in \cite{lerda}, when doing the Kaluza-Klein reduction,
one should be careful with the background expectation values of the
supergravity fields like the volume of the 4-torus since the stability
of the D-brane system depends crucially on the value of this
parameter. Therefore, we will separate the v.e.v. of the bulk fields
from their fluctuations. 
We will not do the complete reduction since, from the boundary state
analysis, we already know that only a subset of the fields will be
active. 
Anticipating on the form of the solution, from the internal part of
the metric, we keep only the fluctuations of the four scalars
associated to its diagonal 
components out of the 58 parameters which parameterize the moduli space
of metrics on K3~: 
\begin{eqnarray}
g_{\mu\mu} = \ex{2\eta_{\mu}}, \qquad \mu=6, \ldots, 9. \nonumber
\end{eqnarray}
The angular deformations of the 4-torus and the moduli associated to
the sixteen fixed points of the orbifold will not be considered. 
The untwisted sector can be obtained using the standard Kaluza-Klein
reduction on a 4-torus and keeping only the states invariant under the
orbifold projection. The only slightly subtle point is that the
${\mathbf Z}_2$ identification halves the volume of compactification.
To write the twisted and Chern-Simmons contributions to the action, we
use the expansions (\ref{expansion}) of the NS-NS and R-R fields on 
the harmonic forms of the orbifold that we normalize in such a way
that 
\begin{eqnarray}
\int_{\rm K3} \tilde{\omega}^I_2
\wedge^{\star_4}\!\tilde{\omega}^J_2=(2\pi\sqrt{\alpha^\prime})^4\delta^{IJ}/2.
\nonumber 
\end{eqnarray}
Including also the untwisted R-R 2-form $C_{(2)}$
(even if we know that the branes system is not charged under this
field and, as we will see in the next section, it can be consistently
put to zero), the reduction of (\ref{action10d}) we need reads
\begin{eqnarray}
{\cal S}^{B,\,\sigma}_{6} &=&
\frac{1}{2\kappa^2_{10}}  \frac{V}{2} 
\Bigg[ \int d^6x \;
\sqrt{-{\det}_6{g}}\; \ex{-2\varphi} R + \int 4\ex{-2\varphi}\; 
d\varphi \wedge^{\star_6}\!{d\varphi}  \nonumber \\ 
&& ~~~~~~~~~~~~~~~~~~~~~~~~~~~~~~~~~~~~~~~~~~~\left. 
- \;\ex{-2\varphi} \; 
d\eta_a \wedge^{\star_6}\!{d\eta_a} - 
\frac{1}{2}{F}_{(3)}\wedge^{\star_6}\!{{F}_{(3)}} \right] \nonumber \\
&& - \frac{(2\pi\sqrt{\alpha^\prime})^4}{4\kappa^2_{10}} 
\int \frac{1}{2} \left[
\ex{-2\phi}\; {\cal H}^I_{(1)} \wedge^{\star_6}\!{\cal H}^I_{(1)} + 
{\tilde{{\cal G}}_{(3)}}^{I} \wedge^{\star_6}\!{\tilde{{\cal G}}_{(3)}}^{I} + 
{\cal A}^I_{(2)} \wedge {\cal H}^I_{(1)} \wedge F_{(3)} \right].
\label{untwisted} 
\end{eqnarray}
in the string frame. We have defined the fluctuations of the
six-dimensional dilaton as 
$$\varphi \equiv \phi - \frac{1}{4}\log({\det}_4{g}) = \phi -
\frac{1}{2}\sum_{a=6}^9 \eta_a$$ 
and the volume $V \equiv \prod_{a=6}^9 (2\pi 
R_a)$ where the $R_a$ are the radii of the 4-torus. In the following,
we will forget the $6$ in $\ ^{\star_6}$ since all bulk actions
and equations will be six-dimensional. 
The Weyl rescaling 
\begin{eqnarray}
g_{\mu\nu} = e^{\varphi} G_{\mu\nu},  \nonumber
\end{eqnarray}
leads us to the action in the six-dimensional Einstein frame~:
\begin{eqnarray}
{\cal S}^B_6 &=& \frac{1}{2 \kappa_{orb}^2} \Bigg[ 
\int d^6x \sqrt{-\det G}\; R - \int \Bigg( d\varphi
\wedge^\star\!{d\varphi} + d\eta_a \wedge^\star\!{d\eta_a} +  
\frac{1}{2}\; \ex{\sum_a{\eta_a}} {F}_{(3)}\wedge^\star\!{{F}_{(3)}}
\nonumber \\ 
&&~~ \left. \left. + \;
\ex{-\sum_a{\eta_a}} {\cal H}^I_{(1)} \wedge^\star\!{\cal H}^I_{(1)} +
\frac{1}{2}
\left( 
{\tilde{\cal G}_{(3)}}^{I} \wedge^\star\!{\tilde{\cal G}_{(3)}}^{I} + 
\sqrt{2}\;{\cal A}^I_{(2)} \wedge {\cal H}^I_{(1)} \wedge F_{(3)}
\right) \right) \right]
\label{SB-Ein}
\end{eqnarray}
where  we have defined the six-dimensional gravitational coupling
constant 
\begin{eqnarray}
\label{con} \kappa_{orb}^2 = \frac{2 \,\kappa_{10}^2}{V} \nonumber
\end{eqnarray}
and we have rescaled  the twisted fields\footnote{In equation
(\ref{SB-Ein}), we have 
suppressed the ' on the redefined fields to lighten the notations.} 
as
\begin{equation}
{{\cal A}'}^I_{(2)} =
\frac{(2\pi\sqrt{\alpha^\prime})^2}{\sqrt{V}}\;{\cal A}^I_{(2)}, \qquad {b'}^I
= \frac{(2\pi\sqrt{\alpha'})^2}{\sqrt{2\;V}}\;{b}^I
\end{equation}
in order to agree with the conventions of \cite{frau}. 
This step is not indispensable
since an other convention would have manifested itself through
different coupling constants in the boundary action and in the
asymptotic values of the fields in the presence of the branes but it
will allow us to blindly use the results of \cite{frau} that we will 
quickly review in the following part before tackling the core subject 
of this article.

\subsection{Boundary actions and the asymptotic conditions}

Using the boundary state which corresponds to a fractional D$p$-brane
and the methods developed in \cite{bs}, the authors of
\cite{frau} found the linear couplings of the bulk fields
to a fractional D-brane and their asymptotic behaviour in the
presence of such source terms. Since our configuration is a superposition
of a fractional D-string and a fractional anti-D-string, we can easily
get from their result the minimally covariantized linear
couplings of the bulk fields to our D-strings pair. 
For a fractional  D$p$-brane which corresponds to a D$p+2$-brane
wrapped on the vanishing 2-cycle $\tilde{\omega}^I$ the effective
action reads~: 
\begin{eqnarray}
{\cal S}_{{\rm D}p}&=&-\frac{M_p}{2}\left\{ 
\int d^{p+1}\sigma
\sqrt{-\det \hat{G}}\; \ex{\frac{(p-1)\varphi}{2}-\sum_a\frac{\eta_a}{2}} 
-\int C_{(p+1)}\right\} \nonumber \\
&-&\frac{M_p \sqrt{V}}{4 \pi^2\alpha'}\left\{ 
\sqrt{2} \int d^{p+1}\sigma
\sqrt{-\det \hat{G}}\;
\ex{\frac{(p-1)\varphi}{2}-\sum_a\frac{\eta_a}{2}}b^I   
-\int \left[{\cal A}^I_{(p+1)}+\sqrt{2}b^I \,C_{(p+1)}\right]\right\}.~~~~~~
\label{boundary-action}
\end{eqnarray} 
The coupling $M_p$ is related to the standard D-brane tension~:
\begin{equation}
M_p \equiv \frac{\sqrt{2}T_p}{\sqrt{V}\kappa_{\rm orb}} \nonumber
\end{equation}
and $\hat{G}$ is the two-dimensional induced metric. 
Looking at the boundary state (\ref{boundary-ad}), we see that we get
the  action for an anti-D-brane by changing the signs in front of the
untwisted R-R form $C_{(p+1)}$ and the twisted NS-NS field $b^I$.  
Therefore, using the boundary state (\ref{dd}) and the action
(\ref{boundary-action}), it is easy to derive the action for the
D{/}$\bar{\rm D}$-brane system~:
\begin{eqnarray}
{\cal S}_{{\rm D}1{/}\bar{\rm D}1}&=&- M_1 \int d^{2}\sigma
\sqrt{-\det \hat{G}}\; 
\ex{-\sum_a\frac{\eta_a}{2}}  
+\frac{M_1 \sqrt{V}}{4 \pi^2\alpha'} 
\Bigg[ 
\int 
\left(
{\cal A}^1_{(2)}+{\cal A}^2_{(2)}
\right) 
\nonumber \\
&& \left.
- \sqrt{2} \int d^{2}\sigma
\sqrt{-\det \hat{G}}\;
\ex{-\sum_a\frac{\eta_a}{2}}
\left(b^1 - b^2\right)  
\right].
\label{dd-action}
\end{eqnarray}
Doing similar calculations for the non-BPS D2-brane (\ref{nbps})
stretched between the 
two fixed points ${\mathbf x}_1$, ${\mathbf x}_2$, leads to~:
\begin{eqnarray}
{\cal S}_{{\rm non-BPS~D}2}&=&- \frac{M_1 R_9}{\sqrt{2\alpha^\prime}} \int
d^{2}\sigma 
\sqrt{-\det \hat{G}}\; 
\ex{\eta_9-\sum_a\frac{\eta_a}{2}}  
+\frac{M_1 \sqrt{V}}{4 \pi^2\alpha'} 
\int 
\left(
{\cal A}^1_{(2)}+{\cal A}^2_{(2)}
\right)
\label{nbps-action}
\end{eqnarray}
In this formula, $\hat{G}$ is also the two-dimensional induced metric on
the non-compact longitudinal space of the D2-brane. 

One can use at least two different methods to determine a supergravity
solution. One is to solve the equations of motion obtained by varying
the sum of the bulk and boundary actions. The other one is to
determine the fields which solve the homogeneous equations
which follow from (\ref{SB-Ein}) and have boundary conditions
dictated by the  
presence of the brane. For the fractional D$p$-brane, this asymptotic
behaviour has also been calculated in \cite{frau}. Since we will adopt
the second procedure in the next section, let us first recall their
result and then, extend it to the non-BPS D-brane and to the
D-string{/}anti-D-string configuration; 
at infinity, in the presence of a fractional D$p$-brane, the metric, the
scalars $\eta_a$, the dilaton and the 
untwisted R-R field have expansions which begin as 
\begin{eqnarray}
&&G_{\alpha\beta} \sim \left(1 + \frac{(p-3)Q_p}{8r^{3-p}}\right) \eta_{\alpha\beta}, \qquad G_{ij} \sim \left(1 +
\frac{(p+1)Q_p}{8 r^{3-p}}\right) \delta_{ij}, \nonumber\\
&&\eta_a \sim \frac{Q_p}{8 r^{3-p}}, \qquad \varphi \sim
\frac{(1-p)Q_p}{8r^{3-p}}, \qquad C_{(p+1)} \sim - \frac{Q_p}{2r^{3-p}}
\end{eqnarray}
while the twisted fields behave as 
\begin{eqnarray}
b^I \sim - \frac{Q_p\sqrt{V}}{4\sqrt{2}\pi^2\alpha^{\prime}r^{3-p}}, \qquad {\cal
A}^I_{(p+1)} \sim - \frac{Q_p\sqrt{V}}{4\pi^2\alpha^{\prime} r^{3-p}}
\end{eqnarray}
where
\begin{eqnarray}
Q_p = \frac{2M_p\kappa^2_{orb}}{(3-p)\Omega_{4-p}}. \nonumber
\end{eqnarray}
From the  boundary states (\ref{nbps}) and 
(\ref{dd}) we can derive similar results for the 
non-BPS D2-brane and for the D-string{/}anti-D-string
configuration. For the non-BPS D2-brane, one gets
\begin{eqnarray}
&&G_{\alpha\beta}\sim\left(1-\frac{Q_1}{2r^2}\frac{R_9}{\sqrt{2\alpha^\prime}}\right)\eta_{\alpha\beta}, 
\qquad
G_{ij}\sim\left(1+\frac{Q_1}{2r^2}\frac{R_9}{\sqrt{2\alpha^\prime}}
\right)
\delta_{ij}, \nonumber \\
&&\eta_a\sim\frac{Q_1}{4r^2}\frac{R_9}{\sqrt{2\alpha^\prime}},  \qquad \mbox{for
}a=6, \ldots, 8,\qquad \eta_9\sim
-\frac{Q_1}{4r^2}\frac{R_9}{\sqrt{2\alpha^\prime}} \nonumber\\   
&&{\cal A}^I_{(2)} \sim - \frac{Q_1\sqrt{V}}{4 \pi^2\alpha^{\prime}
r^2}, \qquad \mbox{for }I=1, 2, 
\label{nbps-asymp}
\end{eqnarray}
and, for the D-string{/}$\bar{\rm D}$-string, 
\begin{eqnarray}
&&G_{\alpha\beta} \sim \left(1 - \frac{Q_1}{2r^2}\right)
\eta_{\alpha\beta}, \qquad G_{ij} \sim \left(1 + 
\frac{Q_1}{2 r^2}\right) \delta_{ij}, \qquad \eta_a \sim
\frac{Q_1}{4r^2}, \nonumber\\ 
&&{\cal A}^I_{(2)} \sim - \frac{Q_1\sqrt{V}}{4 \pi^2\alpha^{\prime}
r^2}, 
\qquad 
b^I \sim - (-1)^I \frac{Q_1\sqrt{V}}{4\sqrt{2}\pi^2\alpha^{\prime} r^2},
\qquad \mbox{for }I=1, 2.
\label{dd-asymp} 
\end{eqnarray}
In the next section, we will use these results to discuss the no-force
condition at leading order in $Q_1$ and solve the equations of motion
with the boundary conditions (\ref{dd-asymp}).

%%%%%%%%%%%%%%%%%%%%%%%%%%%%%%%%%%%%%%%%%%%%%%%%%%%%%%%%%%%%%%%%%%%%%%%%%%
\section{Supergravity description of the D-string{/}anti-D-string
configuration} 

\subsection{No-force condition and critical radii}

One key ingredient in the construction of supergravity solutions which
describe BPS D$p$-branes is that these branes do not exercise any
force on each other. This property, which follows from the
presence of supersymmetry, allows to construct a bound state of an
arbitrary number of D-branes and then to obtain a solution with small
curvature. Therefore, $\alpha^\prime$ curvature corrections to the
supergravity action can be safely neglected. 
This no-force condition is a desirable feature for any type of branes,
including the non-BPS ones, in order to find a reliable classical
solution of the low energy effective action of a string theory. 
In the supergravity picture, such a constraint can be tested by
calculating the potential a non-BPS 
brane probe feels in the geometry created by a bound state of similar
branes. 

We can already investigate this no-force constraint for
the asymptotic part of the fields we have derived in the last
section. To do this, we insert these fields into the action of a brane
probe 
and evaluate the potential. For the non-BPS D2-brane, one gets~:
\begin{equation}
{\cal S}_{\rm probe} = - M_1 \int d^2\sigma
\left[\frac{R_9}{\sqrt{2
\alpha^\prime}}-\frac{R^2_9}{2
\alpha^\prime}\frac{Q_1}{r^2}+\frac{2 V}{(4\pi^2\alpha^\prime)^2} 
\frac{Q_1}{r^2}\right]. 
\end{equation}
Therefore, the potential is constant when 
$\alpha^\prime R_9 = 4 R_6 R_7 R_8$.  When all the radii are critical,
this relation is verified and we recover the one-loop no-force
condition.  However, we also see that, when  $\alpha^\prime R_9 < 4 R_6
R_7 R_8$, which is always the case for a {\it stable} non-BPS D2-brane
configuration with at least one non critical radius, the force is
repulsive.  Therefore, it is impossible to bring at no cost these
non-BPS D2-branes from infinity in order to form a macroscopic bound
state whose supergravity description will be reliable. Finally, if we
want to discuss only configurations which are stable
in the string theory description, the only possibility is for critical
radii. Its T-dual counterpart has been analyzed in \cite{lerda}, where
they found that the no-force condition is broken by higher loops
string corrections and the force is always repulsive.

The same analysis for the D-string{/}$\bar{\rm D}$-string
configuration gives a potential
\begin{equation}
{\cal S}_{\rm probe} = - M_1 \int d^2\sigma
\left[1-\frac{Q_1}{r^2}+\frac{2V}{(4\pi^2\alpha^\prime)^2}
\frac{Q_1}{r^2}-\frac{2V}{(4\pi^2\alpha^\prime)^2}
\frac{Q_1}{r^2}
 \right]  = - M_1\int d^2\sigma \left[1-\frac{Q_1}{r^2}\right] 
\label{dd-potasymp}
\end{equation}
which does not depend on the moduli of the four-torus. Actually, one
sees that, in each twisted sector, the NS-NS and 
R-R contributions cancel each other and 
only remains the attractive potential associated to the untwisted
NS-NS sector. This result is consistent with the analysis of
\cite{lambert}  
where they found that two D{/}\={D}-branes systems always attract
each other. Despite being better than the repulsive force found in
\cite{lerda} which forbids them to construct a macroscopic bound state
of non-BPS branes, one may worry about the absence of a no-force
condition. However, so far, we have only considered the long distance
leading behaviour of the fields in the presence of the D{/}\={D}
pair and it is possible that at shorter distances other phenomenons
occur. For instance, one can imagine that the force vanishes at a
finite distance. 
But, to investigate this question, it is necessary to
go beyond the leading linear behaviour we have discussed so far and to
find the complete solution.

\subsection{The solution}

We will now solve the equations of motion which follow from the action
(\ref{SB-Ein}) with the boundary conditions (\ref{dd-asymp}). To get a
reliable solution at weak string coupling, we should assume that a
macroscopic bound  
state made of $N$ D/\=D pairs can be constructed. Since these D-branes
interact among them in a non trivial way, it is likely that the
couplings of the bound state to the supergravity fields, specially to
the NS-NS sector, are not just
given by the sum of $N$ D/\=D sources (\ref{dd-action}). Actually, in
the appendix, we provide the solution for arbitrary boundary
conditions of the bulk fields, the only hypothesis being that the
bound state excites the same supergravity fields as its
components. However, for simplicity and since we don't know the actual
couplings of our interacting system, we will limit our discussion in
this section to the debatable hypothesis that they are given by the
naive sum of $N$ boundary actions (\ref{dd-action}) and we will
comment on this assumption at the end.

First, since we do not have any source term for the
dilaton and for the untwisted R-R field $C_{(2)}$ and since we saw
that the leading contributions to their asymptotic behaviour 
cancel, we choose to put them to zero. The dilaton equation, which is
simply 
\begin{equation}
d^\star\!d\varphi = 0,
\end{equation}
will be trivially verified. Moreover, it is also easy to see that the
equation for the untwisted R-R field is satisfied for an ansatz which
preserves the symmetry of the problem, namely identical  
twisted R-R potentials and opposite twisted NS-NS scalars.

\noindent The other equations of motion read~:
\begin{eqnarray}
d^\star d\eta_a + \frac{1}{2} \sum_{I=1, 2} \ex{-\sum_a
\eta_a} 
{\cal H}^I_{(1)} \wedge^\star\!{\cal H}^I_{(1)} &=& 0 
\label{eta}
\end{eqnarray}
for the moduli $\eta_a$ ($a=1, \ldots, 4$), 
\begin{equation}
d^\star(\ex{-\sum_a \eta_a}{\cal H}^I_{(1)}) = 0,
\qquad I=1, 2 
\label{bi}
\end{equation}
for the NS-NS twisted scalars $b^I$,
\begin{equation}
d^\star\tilde{\cal G}^I_{(3)}  = 0, \qquad I=1, 2
\label{ai2}
\end{equation}
for the R-R twisted ${\cal A}^I_{(2)}$ and
\begin{eqnarray}
R_{\mu\nu} - \partial_\mu \eta_a \partial_\nu \eta_a - \;
\ex{-\sum_a 
\eta_a} \partial_\mu b^I \partial_\nu b^I 
- \frac{1}{4}\; \tilde{\cal G}^I_{(3)\; \rho
\sigma\mu} \tilde{\cal G}_{(3)\: \nu}^{I\; \rho\sigma} = 0
\label{einstein}
\end{eqnarray}
for the metric. We introduce the following ansatz which depends only
on the radial coordinate and is compatible with the symmetries of the
source~:  
\begin{eqnarray}
&&ds^2 = B^2(r) (-(dx^0)^2 +(dx^1)^2) + F^2(r) (dr^2 + r^2
d\Omega^2_3), ~~~~ r^2=(x^2)^2 + \ldots + (x^5)^2,  \nonumber
\\
&&\tilde{\cal G}^I_{(3)} = d{A}(r) \wedge
dx^0 \wedge dx^1 - 
\ ^\star (d{A}(r) 
\wedge dx^0 \wedge dx^1), \nonumber\\
&&b^I(r)=(-1)^I b(r)~~~~\mbox{and}~~~~ \eta_a(r) = \eta(r).
\label{ansatz}
\end{eqnarray}
After some calculations which are detailed in the appendix, one
arrives at the solution. The fields can be separated into two groups,
reflecting the splitting of the equations into two almost independent
sets.  The scalars $\eta$ and the NS-NS twisted fields $b$ read~:
\begin{eqnarray}
&&\eta = -\frac{1}{2} \ln
\left[\frac{\cosh \Phi}{\sqrt{1+\alpha^2}}\right], \qquad
b=-\frac{1}{\sqrt{2}}\left[\alpha+\sqrt{1+\alpha^2} 
\tanh\Phi \right] 
\label{b-sol}
\end{eqnarray}
where we 
have defined the parameter $\alpha\equiv\sqrt{V_c/2V}$ which depends
on the critical volume $V_c \equiv \prod_a(2\pi R^c_a) = 2(2 \pi^2
\alpha^\prime)^2$ and the functions of the radial coordinates~:
\begin{eqnarray}
&&\Phi \equiv \frac{\sqrt{1+\alpha^2}\;Qy}{2\alpha}-{\rm arc}\sinh\alpha~~~\mbox{and}~~~y(r)\equiv  \frac{2\sqrt{3}}{Q}\; {\rm arctanh} \left(
\frac{Q}{2\sqrt{3}\;r^2}\right).
\label{funct}
\end{eqnarray}
The form of the metric and of the twisted R-R fields depends on the
parameter $\alpha$. For a
volume strictly larger than $V_c/2$, we get~:
\begin{eqnarray}
&&A=\sqrt{1-\alpha^2}\; {\rm cotan}\;\Theta -
\alpha,~~~~B^{2}=\frac{\sqrt{1-\alpha^2}}{\sin\Theta},~~~~
F^2 = f_+ f_- B^{-2},~~~~  
\label{metric1}
\end{eqnarray}
where we have introduced~:
\begin{eqnarray}
\Theta \equiv \frac{\sqrt{1-\alpha^2}\;Q
y}{2\alpha}+\arccos \alpha~~~~\mbox{and}~~~~f_\pm(r) \equiv 1 \pm
\frac{Q}{2\sqrt{3}\;r^2}. 
\end{eqnarray}
For a volume smaller than $V_c/2$, one must replace in
(\ref{metric1}) the trigonometric functions by their hyperbolic
counterparts and change $\sqrt{1-\alpha^2}$ to
$\sqrt{\alpha^2-1}$~:
\begin{eqnarray}
&&A=\sqrt{\alpha^2-1}\; {\rm cotanh}\;
\tilde{\Theta} -
\alpha,~~~~B^{2}=\frac{\sqrt{\alpha^2-1}}{\sinh\tilde{\Theta}},~~~~F^2
= f_+ f_- B^{-2}, 
\nonumber \\
&&\tilde{\Theta} \equiv \frac{\sqrt{\alpha^2-1}\;Q
y}{2\alpha}+{\rm arccosh}\; \alpha .
\label{metric2}
\end{eqnarray}
Finally, the solution at $V_c/2$ can be obtained as
a degenerate limit of these two cases~:
\begin{eqnarray}
&&A = -\frac{Qy}{2+Qy},~~~~B^2 = \frac{2}{2+Qy},~~~~F^2 = f_+ f_- B^{-2}.
\label{metric3}
\end{eqnarray}
Let us discuss the properties of these solutions. 
First, we can notice that the functions $f_\pm$ and $y$ do not depend
on the volume~: the reason lies in a cancellation between the
contributions of the twisted R-R and NS-NS sectors. The second 
observation we can make is that the ``transition'' between solutions
(\ref{metric1}) and (\ref{metric2})
does not occur at the critical volume, contrary to one can have
naively expected. However, we have to remind the reader that this
result is 
related to our working hypothesis, namely that the couplings of the
bound state are given by the naive sum of its components. The third
point we would like to emphasize is that 
the solution depends on the radii only through the volume of the
four-torus. This means that it can not capture the detailed structure
of the stability domain of the non-BPS system. For
instance, when $V>V_c$, the same supergravity solution will describe
configurations 
which are stable and configurations which are unstable from the string
theory point-of-view. Therefore, we don't expect to see transitions
between stable and unstable configurations in these solutions. This
remark can also explain why the critical volume does not seem to play any
special role. 

Then, one can calculate the scalar curvature
\begin{equation}
R= \left(1+\frac{1}{\alpha^2}\right)\frac{B^2}{(r^2 f_+f_-)^3}
\label{curvature}
\end{equation}
and  investigate the presence of possible
singularities.  

In the case $V>V_c/2$, the
metric and the twisted R-R fields have branch cut singularities at
radii defined by 
$$y_n \equiv y(r_n) = \frac{2 \alpha (n \pi - \arccos
\alpha)}{Q\sqrt{1-\alpha^2}}, \qquad n=1, 2, \ldots$$ where the
longitudinal components of the metric and the curvature
(\ref{curvature}) diverge. Moreover, the first of these singularities
deserves the name of a ``repulson'' since the Newtonian gravitational
force becomes repulsive in the region $]r_1,
r_r[$ where the radius $r_r$ defined by 
\begin{eqnarray}
y(r_r) = \frac{2 \alpha (\pi/2 - \arccos
\alpha)}{Q\sqrt{1-\alpha^2}}. \nonumber
\end{eqnarray}

When the volume becomes equal to or larger than $V_c/2$,
these branch cut singularities disappear and only remains a
singularity located at
$y(r_0)=\infty$ or, using (\ref{funct}), at a radius
$$r^2_0={Q/2\sqrt{3}}.$$
This singularity is rather different from the other singularities
obtained for $V>V_c/2$ since now, the longitudinal components of the
metric vanish as one reaches it. The curvature divergence is due to 
the presence of the function $f_-$ in the scalar curvature. Finally,
this singularity is always attractive. 

These naked singularities make the solutions unacceptable according to
the cosmic censorship. However, we know D-brane configurations which
also give rise to singularities but, in some cases, this pathological
behaviour can be solved. An example is the enhan\c{c}on mechanism
\cite{john} which solves the singularity of the metric which
corresponds to 
fractional D-branes. In the next subsection, we will argue  for a
similar resolution of the singular solutions generated by D/\={D}
pairs.

\subsection{D-brane probe and the enhan\c{c}on mechanism}

To fix the ideas, we will consider only the case $\alpha>1$ but the other
two cases are completely similar as we will comment at the end.
To understand what is happening when, falling from infinity, 
we are approaching the first of these singularities, we consider a
probe D-string{/}\={D}-string pair slowly moving
in the geometry defined by (\ref{b-sol}) and
(\ref{metric1}). In the static gauge, its space-time coordinates are
$\zeta^\alpha=x^\alpha$ and $\zeta^i=\zeta^i(x^0)$. 
Its velocity is given by $v_i = \dot{\zeta}^i$. Then, using
(\ref{b-sol}) and 
(\ref{metric1}) in the boundary action (\ref{dd-action}), and
expanding in powers of $v^2$, we get
\begin{eqnarray}
{\cal S}_{\rm probe} &=& - M_1 \int d^2\sigma
\left[ 
1-
\frac{\sqrt{1-\alpha^2}\left(\cos\Theta + \sinh\Phi\right)}{\alpha\sin\Theta}
+\frac{f_+ f_- \sin\Theta\; \sinh 
\Phi \;v^2}{2\alpha \sqrt{1-\alpha^2}}  + {\cal O}(v^4)
\right] \nonumber\\
\label{kinprobe}
\end{eqnarray}
The $v$ independent term corresponds to the potential felt by the probe
in the supergravity 
background. Expanding this term when $r\rightarrow \infty$, one
obviously 
recovers the attractive force we had already observed at infinity
(\ref{dd-potasymp}). One can also verify that this force remains
attractive in the whole region $]r_1, +\infty[$\footnote{Since the
force is attractive, the velocity and the energy 
of the probe can not be kept constant at the same time. Therefore, one
may worry about the use of a non-relativistic approximation for the
probe action. Actually, one can repeat the above analysis for a
relativistic probe, like in \cite{berglund}, and verify that the
conclusion remains unchanged.}.

We also observe that the sign of the $v^2$ dependent part changes at a
radius $r_e>r_1$ defined by $\Phi=0$~:
\begin{equation}
y(r_e) = \frac{2\alpha\; {\rm arcsinh}\;
\alpha}{Q\sqrt{1+\alpha^2}}.
\end{equation}
The D{/}\={D}-string pair falling from infinity becomes 
tensionless before reaching the singularity. The vanishing 
of the effective tension can be traced back to the non trivial
contributions of 
the twisted NS-NS fields $b^1$ and $b^2$ to the Dirac-Born-Infeld part
of the action.  Indeed, at infinity, the fluctuations of these fields
vanish (we remind the reader that the {\it vev} of the twisted NS-NS fields
are $1/2$ in this compactification). However, when we approaching the
core, these fluctuations decrease and exactly compensate their
background value at $r_e$.

The existence of a tensionless sphere is reminiscent of the situation
studied in \cite{john} and we 
can suggest a similar resolution of the singularity; the radius $r_e$
at which the probe 
becomes  
tensionless was called the enhan\c{c}on radius in \cite{john}~: the
reason was that, at this radius, new massless bulk states appeared and
the gauge group symmetry was enhanced. 

To see if a similar enhan\c{c}on mechanism can  also be argued for 
solving the singularity of our solution,  it is convenient
to discuss the theory from a type IIA on $T^4/{\mathbf Z}_2$
point-of-view. Therefore, we 
T-dualize our solution along a direction transverse to $K3$ 
and to the D-strings.  The dual type IIA string theory on $T^4/{\mathbf
Z}_2$ 
has a gauge group $U(1)^{24}$, with $16$ $U(1)$ associated to the R-R
twisted 1-forms ${\cal A}^I_{(1)}$, which come from the reduction of
the R-R 3-form $C_{(3)}$ on the vanishing supersymmetric 2-cycles
$\tilde{\omega}^I_2$, $6$ $U(1)$ associated to the R-R untwisted 1-forms
${C}^i_{(1)}$, obtained by reducing $C_{(3)}$ on the 2-cycles of the
torus and $2$ $U(1)$ whose gauge potentials are the R-R 1-form
$C_{(1)}$ and the six-dimensional Hodge-dual of $C_{(3)}$. 
After T-duality, the D-string/\={D}-string system becomes a  
D2-brane/\={D}2-brane configuration, which is magnetically charged under
two twisted R-R 1-forms ${\cal A}^1_{(1)}$ and ${\cal A}^2_{(1)}$. Therefore, our
D2-brane/\={D}2-brane system appears as a non-BPS monopole of the
gauge group $U(1)_1\otimes U(1)_2 \subset U(1)^{24}$. 
The twisted NS-NS fields remain the same. 

At the enhan\c{c}on radius, one can notice that an enhancement of these
gauge 
groups occurs. This enhancement is due to fractional D0-branes,
electrically charged under ${\cal A}^1_{(1)}$ or ${\cal A}^2_{(1)}$,
which become massless at $r_e$. To see this, consider the action for
such fractional D0-branes, {\it i.e.} eqn.(\ref{boundary-action})
with $p=0$; 
expanding the Dirac-Born-Infeld part of the action to lowest order in
the velocity, we obtain the following kinetic term~:
\begin{eqnarray}
{\cal S}_{\rm kin.} = - \frac{M_0}{2} \int d\sigma \; \frac{f_+ f_- \sin
\Theta \sinh \Phi}{2\alpha \sqrt{1-\alpha^2}} \; v^2
\end{eqnarray}
which is obviously proportional to the one of the
D-string/\={D}-string probe (\ref{kinprobe}).  Again, the vanishing
of the 
effective tension is completely determined by the fluctuations of the
twisted NS-NS fields, $b^1$ and $b^2$, and occurs when these fluctuations
compensate their background values. Therefore, the masses of the
(anti-)fractional D-particles located at the fixed points ${\mathbf
x}_I, I=1, 2$ vanish at $r_e$ and, similarly to
\cite{john}, we can conclude that, at $r_e$, stringy phenomena such as
the enhancement of the gauge symmetry  play a role and that the
solution inside this special radius should not be taken seriously. 
Our supergravity solution should only be taken as physical down to the
sphere of radius $r_e$ and the $N$ D{/}\={D}-string pairs live on this
sphere.   

Using the S-duality  which relates type IIA string theory
compactified on $T^4/{\mathbf Z}_2$ and heterotic string on $T^4$, we
can also give an heterotic interpretation of this gauge symmetry
enhancement. The precise mapping of the moduli and of the gauge fields
given in \cite{kiritsis} tells us that the 16 twisted NS-NS moduli $b^I$
are identified to Wilson lines values on one of the four compact
directions of the 16 $U(1)$ fields, $A^K$, which form the Cartan torus
of the broken $SO(32)$ group on the heterotic side. 
Using the mass formula for
heterotic states 
given for example in \cite{gaberdiel2}, 
\begin{eqnarray*}
&&\frac{1}{4}m^2_h = P^2_L + 2(N_L - 1), \\
&&P_L = (V^K + A^K_a w^a, \frac{p_a}{2 R_a} + w^a R_a), \qquad p_a =
n_a + B_{ab}w^b - V^K A^{K}_a - A^{K}_a A^K_b w^b/2, 
\end{eqnarray*}
it is easy to verify that, when one of these
values vanishes, new massless states appear and the associated $U(1)$
gauge group is enhanced to $SU(2)$.  For instance, when the first two
twisted NS-NS 
moduli are shifted by $-1/2$, these additional massless gauge
bosons are given by $N_L=0$, $w^a=0, p_a=0$ for all $a$ and charges
$V=\pm(1,\pm 1, 0^{14})$. We remind the reader that the sixteen
twisted $U(1)$ charges in the type IIA picture 
correspond to symmetric and antisymmetric combinations of the
$(2n+1)$'st and $(2n+2)$'nd $U(1)$ charges in the heterotic
description. 

Since this mechanism only depends on the twisted NS-NS fields, whose
form is independent of the parameter $\alpha$, it will also work for
$V \leq V_c/2$. The resolution of the singularities for any volume may
seem strange since the string theory analysis 
predicts a tachyonic instability when $V < V_c$. However, as we have
already argued, our supergravity description can not capture the
transitions between stable and instable configurations, as it is shown
by the cases $V>V_c$ which cover both situations. We have also to keep in
mind the two hypothesis we have made to find this solution~: first, we
have neglected the dipole effects due to the separation 
of the D-branes along the compact directions. Taking into account this
effect would probably lead to a solution which depends on the 
different radii not just through the volume. Therefore, it would be
possible to investigate its behaviour under variations of the radii.

The second hypothesis was to take the sum of $N$ D/\=D pair
actions as a source.  It is instructing to relax this
hypothesis and see if something special happens for specific values of
the couplings of the bound state to the bulk fields. The total charges
of the system under the R-R twisted fields are always given by the sum
of $N$ charges and their asymptotic values are governed by $A_\infty =
Q\sqrt{V/2V_c}$ but the couplings to the NS-NS fields can be 
modified. Let us consider generic asymptotic values for the metric,
$\eta_a$ and $b^I$ (respectively called $a$, $\eta_\infty$ and
$b_\infty$ in the appendix.). 
First,  the nature of the singularities is governed by $\gamma \equiv a^2 -
A^2_\infty$. When  $\gamma < 0$, we have always branch cut singularities
while when $\gamma \geq 0$, the parameter  $c = \sqrt{(\gamma + 4
\eta_\infty^2 + 2 b_\infty^2)/6}$ introduced in the
appendix is always a non zero real number and there is only one
singularity, located at $y(r)=\infty$.
 The radius of the enhan\c{c}on sphere is governed by the parameters $a$ and
$b_\infty$. At fixed $a$, its position, $y_e$, decreases
from $\infty$ to $0$ (the corresponding radius increases) when
$b_\infty$ goes from $0$ to $\infty$. In particular, 
the case $b_\infty=0$ reproduces the T-dual version of the result of 
\cite{lerda}. Therefore, for $\gamma \geq 0$, the singularity is always
inside the enhan\c{c}on sphere and coincides with it at the limit
$b_\infty \rightarrow 0$.  

On the other hand, when $\gamma < 0$, for
given values of $a$ and $\eta_\infty$, at small enough 
$b_\infty$, the first singularity will be outside the enhan\c{c}on
sphere. In this case, its naked singularities can not be
resolved by the enhan\c{c}on mechanism. Moreover, one can see that it
is impossible to create a horizon just by playing on the values of the
three parameters while staying in this situation. The non-BPS D-brane
phase of \cite{lerda} always corresponds to this latter case.

%%%%%%%%%%%%%%%%%%%%%%%%%%%%%%%%%%%%%%%%%%%%%%%%%%%%%%%%%%%%%%%%%%%%%%%%%%
\section{\bf Discussion}

In this paper, we have found the supergravity description of
D-strings{/}\={D}-strings  located at opposite
fixed points of the orbifold IIB$/({T^4}/{\mathbf Z}_2)$. These
D-pairs feel an attractive force, while the force was 
repulsive for the non-BPS D-branes. The other important difference with
the non-BPS D-brane is that the D/\=D pair is a source for
twisted NS-NS scalars. Our
solutions display singularities but we have observed that a probe in
this background becomes tensionless before reaching the first
singularity, suggesting a enhan\c{c}on mechanism similar to the one
advocated in \cite{john}. Actually, we have shown that, on the
tensionless sphere, new massless gauge fields appear and that the
gauge group is enhanced. 
However, some questions are still open; indeed, it is impossible to
see the effect of the tachyonic instability predicted by string theory
in our solutions. This problem can be related to our working
hypothesis. Indeed, the transitions between different D-branes phases
depend on the value of each compactification radius, something that
our solution does not capture. 

Finally, let us speculate on how supergravity  would select the
``correct'' solution. First, we expect that it will 
have no pathology and therefore that its singularities have been
resolved. Then, the well-behaved solutions differ by their couplings
to the NS-NS fields, which, due to interactions between the
D-branes, can not be fixed easily. However, one could expect that the
ground state would correspond to the solution with the smallest ADM
mass.  

Such solutions can have applications in non-supersymmetric
versions of the AdS/CFT correspondence \cite{maldacena, malrv}. In
this context of dual 
description of D-brane{/}anti-D-brane pairs, let
us conclude with some comments on a related but probably harder
question which has been raised by Mukhi and Suryanarayana
\cite{mukhi2}.  In the type IIA picture, they have considered
adjacent D4-branes{/}\={D}4-branes pairs stretched between two
relatively rotated NS5-branes.  They said that they expected a repulsive
force between the endpoints of the adjacent brane and anti-brane but,
since the D-branes can not be separated without being stretched, they
argued that it should exist a critical value of the distance between
the branes endpoints such that the repulsive force is compensated by
the tension.  In the T-dual type IIB language, this system
is interpreted as fractional D3-branes{/}\={D}3-branes pairs
sitting at the fixed point of a conifold. The 
fractional D3-branes are wrapped D5-branes on one of the two 2-spheres
of the base of the conifold, $T^{1,1}$. Separating the branes in the
type IIA picture may correspond to a deformation of the conifold;
in particular, for symmetric reasons, the wrapped D5-brane and  
anti-D5-brane should be located at antipodal points of the other
2-sphere and the cancellation argued in \cite{mukhi2} should
demonstrate itself as the existence a radius of the sphere for which
the system 
is stable.  However, it is likely that the repulsive force gets higher
loops corrections which are difficult to calculate in string
theory. One can think that, as in \cite{lerda}, the classical
description will tell us something about these corrections but this
question is certainly difficult to solve since it involves dipole
solutions which are 
not well understood so far.

\

%%%%%%%%%%%%%%%%%%%%%%%%%%%%%%%%%%%%%%%%%%%%%%%%%%%%%%%%%%%%%%%%%%%%%%%%%%%%%%%%%%%%%%%%%%%%%%%%%
\noindent{\bf Acknowledgements} 

We would like to thank M. Green for useful suggestions. We are also
grateful to  C. Bachas for illuminating comments and questions.  We
acknowledge PPARC for financial support.  

%This work was supported by a PPARC grant. 

\vskip 1.3cm 
\appendix{\noindent\Large {\bf{Appendix}}} 
\label{appendix} 

\

In this appendix, we solve explicitly for the ansatz (\ref{ansatz})
the equations of motion complemented by the generic boundary
conditions~:
\begin{eqnarray}
B^2 \sim 1 - \frac{a}{r^2}, ~~~~ G^2 \sim  1 + \frac{a}{r^2},~~~~ \eta \sim
\frac{\eta_\infty}{r^2}~~~~
A \sim - \frac{A_\infty}{r^2}, ~~~~\mbox{and}~~~~ b \sim -
\frac{b_\infty}{r^2}. 
\label{gen-asymp}
\end{eqnarray}
Identifying the asymptotic constants as
\begin{eqnarray}
a=\frac{Q}{2},~~~~ \eta_\infty = \frac{Q}{4},~~~~ b_\infty =
\frac{Q}{2}\sqrt{\frac{V}{V_c}}~~~\mbox{and}~~~A_\infty =
\frac{Q}{\sqrt{2}}\sqrt{\frac{V}{V_c}}.
\label{asymptotic}
\end{eqnarray}
we will recover the boundary conditions (\ref{dd-asymp}).
Introducing the field $\xi\equiv 2(\ln B + \ln F)$, the field
equations become 
\begin{eqnarray}
\partial_r\left(r^3 \ex{\xi - 4 \eta} \partial_r b^I \right) = 0 
\label{b-eq}
\end{eqnarray}
for the NS-NS twisted scalars,
\begin{eqnarray}
\partial_r\left(r^3 \ex{\xi} B^{-4} \partial_r A^I \right) = 0 
\label{A-eq}
\end{eqnarray}
for the R-R twisted 2-forms,
\begin{eqnarray}
\partial_r\left(r^3 \ex{\xi} \partial_r \eta \right) + r^3 \ex{\xi - 4
\eta} \left(\partial_r b\right)^2 = 0
\label{eta-eq}
\end{eqnarray}
for the NS-NS untwisted scalars and 
\begin{eqnarray}
&&R^r_{\ r} - \frac{1}{F^2}\left(4(\partial_r \eta)^2 + 2 \ex{-4\eta}(\partial_r b)^2 -
\frac{1}{2 B^{4}}(\partial_r A)^2 \right) = 0, \nonumber \\
&&R^\theta_{\ \theta} - \frac{1}{2 F^2 B^4}(\partial_r A)^2 = 0,\nonumber \\
&&R^\alpha_{\ \alpha} + \frac{1}{2 F^2 B^4}(\partial_r A)^2 = 0, 
\label{R-eq}
\end{eqnarray}
for the radial, transverse and longitudinal components of the Ricci
tensor. 
The first two equations, which govern the dynamics of the twisted
fields,  can be integrated to~:
\begin{eqnarray}
&&\partial_r b = \frac{2b_\infty}{r^3}\;  \ex{-\xi+ 4\eta}, \nonumber \\
&&\partial_r A = \frac{2A_\infty}{r^3} B^4 \ex{-\xi}
\label{twis-eq}
\end{eqnarray}
where the boundary constraints (\ref{gen-asymp}) have been used to
determine the constants of  integration. 
Then, one reinjects these values in the formula (\ref{eta-eq}) and
(\ref{R-eq}). Equation (\ref{eta-eq}) becomes~:
\begin{equation}
\partial_r\left(r^3 \ex{\xi} \partial_r \eta \right) + \frac{4 b_\infty^2}{r^3}
\;\ex{-\xi + 4\eta} = 0
\label{eta-eq2}
\end{equation}
that can be integrated to
\begin{equation}
\frac{r^3}{2} (\partial_r \eta)^2 + \frac{b_\infty^2}{r^3}
\ex{-2\xi + 4\eta} = \frac{b_\infty^2 + 2 \eta_\infty^2}{r^3}\; \ex{-2\xi}.
\label{eta-eq3}
\end{equation}
Using equations (\ref{twis-eq}) and (\ref{eta-eq3}) in the
Einstein equations (\ref{R-eq}), we get 
\begin{eqnarray}
&&\left[-\partial_r \!(r^3 \partial_r \!\ln F) + r^3 \partial_r \xi
\partial_r \!\ln F - r^3\partial_r^2 \xi - 2r^3\left((\partial_r\!\ln
B)^2+(\partial_r \!\ln F)^2\right)\right] \nonumber \\
&&~~~~~~~~~~~~~~~~~~~~~~~~~~~~~~~~~~~~~~~~~~~~ -
\frac{8(b_\infty^2+2\eta_\infty^2)}{r^3}\; \ex{-2\xi} +  
\frac{2A_\infty^2}{r^3}B^4 \ex{-2\xi} = 0, 
\label{Rrr-eq}
\\
&&\left[-\partial_r \!(r^3 \ex{\xi} \partial_r \!\ln F) - r^2
\partial_r \ex{\xi}\right] - \frac{2A_\infty^2}{r^3}B^4 
\ex{-\xi} = 0, 
\label{Rtt-eq}
\\ 
&&\left[-\partial_r \!(r^3 \ex{\xi}\partial_r \!\ln B)\right] + \frac{2A_\infty^2}{r^3}B^4 \ex{-\xi} = 0,
\label{Raa-eq}
\end{eqnarray}
where we have also introduced the explicit form of the Ricci
tensor. We see that the equations split into two almost
independent groups which are only connected by the function $\xi$~:
one for  
the six-dimensional metric and the twisted NS-NS fields, and the other
for the untwisted NS-NS scalars and the twisted R-R forms. 

The strategy is to first determine the field $\xi$ and
then solve the other equations separately. Summing the last two
equations, we see that $\xi$ obeys the homogeneous differential
equation 
\begin{equation}
\partial_r (r^5 \partial_r \ex{\xi}) = 0 
\end{equation}
whose solution is of the form
\begin{equation}
\ex{\xi} = f_{+}(r) f_{-}(r)
\end{equation}
with
\begin{equation}
f_{\pm}(r) = 1 \pm \frac{c}{r^2}.
\end{equation}
The constant of integration $c$ will be determined at the end of the
calculation, by 
inserting into equation (\ref{Rrr-eq}) the solutions we will find
below. It is easy to see that the fields $B$ and  
$\eta$ can be expressed in term of the function~:
\begin{equation}
y(r)\equiv \frac{1}{2 c} \ln \left(\frac{f_+(r)}{f_-(r)}\right)
\end{equation}
In term of this new variable, equations (\ref{eta-eq2}) and
(\ref{Raa-eq}) simply read
\begin{equation}
\partial_y^2 \eta + b_\infty^2 \ex{4\eta} = 0
\end{equation}
and
\begin{equation}
2 \partial_y^2 \ln B - A_\infty^2 B^4 = 0
\end{equation}
which are respectively solved by
\begin{equation}
\eta = -\frac{1}{2} \ln \left(\frac{\cosh \left(\sqrt{2}b_\infty
\sqrt{1+\alpha^2}\; y - {\rm arcsinh}\; \alpha
\right)}{\sqrt{1+\alpha^2}}\right)
\end{equation}
and
\begin{equation}
B^2=\frac{\sqrt{\beta^2-1}}{\sinh\left(A_\infty
\sqrt{\beta^2-1}\; y +
{\rm arccosh}\;\beta\right)} 
\end{equation}
where the appropriate  boundary conditions (\ref{gen-asymp}) have been
used and we have defined 
\begin{eqnarray}
\alpha\equiv\frac{\sqrt{2}\eta_\infty}{b_\infty}~~~~~\mbox{and}~~~~~\beta\equiv\frac{a}{A_\infty}.
\end{eqnarray}
Then, to obtain the twisted fields, we integrate (\ref{twis-eq}). The
result is
\begin{eqnarray}
&&b= - \frac{\alpha +
\sqrt{1+\alpha^2} \tanh\left(\sqrt{2}b_\infty
\sqrt{1+\alpha^2}\; y - {\rm arcsinh}\;\alpha\right)}{\sqrt{2}}, \nonumber \\
&&A= - \beta + \sqrt{\beta^2-1}\; {\rm cotanh}\left(A_\infty
\sqrt{\beta^2-1}\; y +
{\rm arccosh}\;\beta\right).
\end{eqnarray}
Finally, inserting these formula into equation (\ref{Rrr-eq}) allows
us to determine the constant 
\begin{equation}
c=\sqrt{\frac{1}{6}\left(a^2+2b^2_\infty+4\eta_\infty^2-A_\infty^2\right)}
\end{equation}
In section 4, we specialize this general result to the boundary
conditions (\ref{asymptotic}).

\vfill
\eject
%%%%%%%%%%%%%%%%%%%%%%%%%%%%%%%%%%%%%%%%%%%%%%%%%%%%%%%%%%%%%%%%%%%%%%%%%%%%%%%%%%%%%%%%
%\setcounter{section}{0}   %  starts Appendix lettering at "A"
%\Appendix{}

%%%%%%%%%%%%%%%%%%%%%%%%%%%%%%%%%%%%%%%%%%%%%%%%%%%%%%%%%%%%%%%%%%%%%%%%%%%%%
%\newpage

%%%%%%%%%%%%%%%%%%%%%%%%%%%%%%%%%%%%%%%%%%%%%%%%%%%%%%%%%%%%%%%%%%%%%%%%%
\end{document}